\documentclass[%
 reprint,%
 amssymb, amsmath,%
 aip,jap,%
]{revtex4-1}

\usepackage{graphicx}
\usepackage{bm}%
\usepackage[colorlinks=true,linkcolor=blue]{hyperref}%
\newcommand{\xax}{$\bf{\hat{x}}$}

\newcommand{\zax}{$\bf{\hat{z}}$}
\expandafter\ifx\csname package@font\endcsname\relax\else
 \expandafter\expandafter
 \expandafter\usepackage
 \expandafter\expandafter
 \expandafter{\csname package@font\endcsname}%
\fi
\begin{document}

\title{Non-volatile spin wave majority gate at the nanoscale}%

\author{O. Zografos}
\email{Odysseas.Zografos@imec.be}
\affiliation{imec, Kapeldreef 75, B-3001 Leuven, Belgium}
\affiliation{KU Leuven, ESAT, B-3001 Leuven, Belgium}
\author{S. Dutta}
\affiliation{Department of Electrical and Computer Engineering, Georgia Institute of Technology, 30332 Atlanta, GA, USA}
\author{M. Manfrini}
\author{A. Vaysset}
\affiliation{imec, Kapeldreef 75, B-3001 Leuven, Belgium}
\author{B. Sor\'{e}e}
\affiliation{imec, Kapeldreef 75, B-3001 Leuven, Belgium}
\affiliation{KU Leuven, ESAT, B-3001 Leuven, Belgium}
\affiliation{Universiteit Antwerpen, Physics Department, B-2020 Antwerpen, Belgium}
\author{A. Naeemi}
\affiliation{Department of Electrical and Computer Engineering, Georgia Institute of Technology, 30332 Atlanta, GA, USA}
\author{\\P. Raghavan}
\affiliation{imec, Kapeldreef 75, B-3001 Leuven, Belgium}
\author{R. Lauwereins}
\affiliation{imec, Kapeldreef 75, B-3001 Leuven, Belgium}
\affiliation{KU Leuven, ESAT, B-3001 Leuven, Belgium}
\author{I. P. Radu}
\affiliation{imec, Kapeldreef 75, B-3001 Leuven, Belgium}

\date{\today}%

\begin{abstract}
A spin wave majority fork-like structure with feature size of 40\,nm, is presented and investigated, through micromagnetic simulations. The structure consists of three merging out-of-plane magnetization spin wave buses and four magneto-electric cells serving as three inputs and an output. The information of the logic signals is encoded in the phase of the transmitted spin waves and subsequently stored as direction of magnetization of the magneto-electric cells upon detection. The minimum dimensions of the structure that produce an operational majority gate are identified. For all input combinations, the detection scheme employed manages to capture the majority phase result of the spin wave interference and ignore all reflection effects induced by the geometry of the structure.
\end{abstract}

\maketitle

The exploration and study of novel non-charge-based logic devices has been a main research focus for over a decade.\cite{Zhirnov2003} The purpose is to identify concepts that can extend the semiconductor industry roadmap beyond the complementary metal oxide semiconductor (CMOS) technology.\cite{Hutchby2002} Since CMOS scaling, dictated by Moore's Law,\cite{Moore1965} will reach its limits,\cite{Zhirnov2003} there is a need for logic components that can operate at high frequencies, be extremely compact and also consume ultra-low power.\cite{nikonov} A variety of magnetic devices have been benchmarked as promising candidates for low power applications.\cite{nikonov} Spin wave devices hold the promise of ultra-low power per computing throughput.\cite{nikonov} Additionally, utilizing spin waves, majority-based logic can be constructed and has been proven to be advantageous for beyond-CMOS technologies.\cite{nanoarch,amaru} These devices have been extensively studied through experiments and micromagnetic simulations at large dimensions (down to tens of microns),\cite{klinger1,klinger2} however the study of spin wave dynamics and interference at the nanoscale are still lacking. 

\begin{figure}[!ht]
	\includegraphics{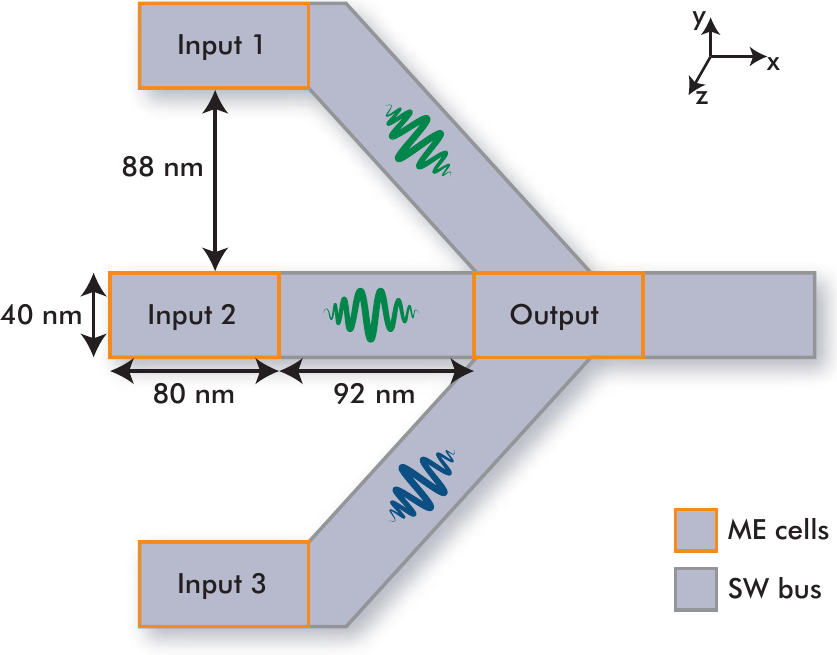}
	\caption{\label{fig:struct} Geometry of the spin wave majority gate. Spin waves are excited by the three input ME cells (Inputs 1,2,3) and the majority result of the spin wave interference is detected by the `Output' ME cell. The spacing between each arm is $S$=88\,nm. }
\end{figure}

In this work, we investigate through micromagnetic simulations, a fork-like spin wave majority structure with feature size of 40\,nm. We aim at designing a nanometer scale structure where excitation of higher-order width modes\cite{klinger2} can be avoided. The proposed design incorporates the advantages of non-volatile data storage in the ME cell, non-reciprocity via a three-phase clocking scheme\cite{sourav1,sourav2} and robustness to thermal fluctuations missing in the earlier prior designs.\cite{klinger1,klinger2,khitun} The structure consists of three merging perpendicular magnetic anisotropy (PMA) spin wave buses and four magneto-electric (ME) cells serving as three inputs and an output. The geometry of the spin wave majority gate is shown in FIG. \ref{fig:struct}, where the spacing between each arm is $S$=88\,nm.

We employed micromagnetic simulations to investigate this structure, using the micromagnetic solver OOMMF.\cite{oommf} The mesh cell size is 2\,nm$\times$2\,nm$\times$12\,nm\cite{supp} and all the PMA spin wave bus regions are extended before and after the ME cell regions with increased damping to allow for magnetization relaxation and avoid edge reflections. Thus, the simulated structure represents an spin wave majority gate arrangement on infinitely long buses. The extended regions of the structure are not shown in FIG. \ref{fig:struct} for ease of representation.

The basic computational block of a spin wave logic device is the ME cell that acts as a spin wave transmitter, detector and also serves as a non-volatile memory element.\cite{sourav1} The ME cells are embedded in the bus and have in-plane magnetization (along $\pm$\xax{}). They are heterostructures consisting of a ferroelectric or piezoelectric material intelayered between two metallic electrodes and a top magnetostrictive ferromagnetic layer. 

We consider a 80\,nm$\times$40\,nm$\times$12\,nm Co$_{60}$Fe$_{40}$/(001) PMN-PT (30\,nm thick) as the ME cell heterostructure (with magnetization saturation M$_{\textrm S}$=800\,kA/m, exchange constant $A$=20\,pJ/m, Gilbert damping $\alpha$=0.027, magnetostrictive coefficient $\lambda$=200\,ppm, Young's modulus $Y$=200\,GPa, and piezoelectric coefficient $d_{31}$=-1000\,pm/V). (001) PMN-PT is chosen as the piezoelectric layer due to its high piezoelectric coefficient while Co$_{60}$Fe$_{40}$ displays a large magnetostrictive coefficient of 2$\cdot$10$^4$,\cite{hunter} and is also compatible with PMN-PT. The spin wave bus material is to be considered a [Co(0.4)/Ni(0.8)]$_{10}$ multilayer (with M$_{\textrm S}$=790\,kA/m, $A$=16\,pJ/m, $\alpha$=0.01, and anisotropy field $H_{K}$=16.78\,kA/m). It is selected as the spin wave bus material due to its inherent interface anisotropy, thus providing a bias-free out-of-plane magnetic configuration. The working principle is based on voltage-controlled strain-induced magnetization switching that excites spin waves and a phase dependent deterministic detection scheme, where information is encoded in the phase of the transmitted spin wave and subsequently stored as direction of magnetization of the ME cell (+\xax{} or -\xax{}).\cite{sourav1,sourav2} 

An applied voltage across the piezoelectric layer causes an isotropic biaxial strain that gets coupled to the top ferromagnet causing an out-of-plane anisotropy. Above a critical strain, the magnetization switches from an in-plane to out-of-plane configuration exciting spin waves with the information encoded in the phase of the waves. Meanwhile, the detector ME cell is held out-of-plane via application of voltage until the spin waves arrive. Upon arrival, the voltage is turned off causing a phase-dependent deterministic switching of the magnetization. 

\begin{figure}[!ht]
	\includegraphics{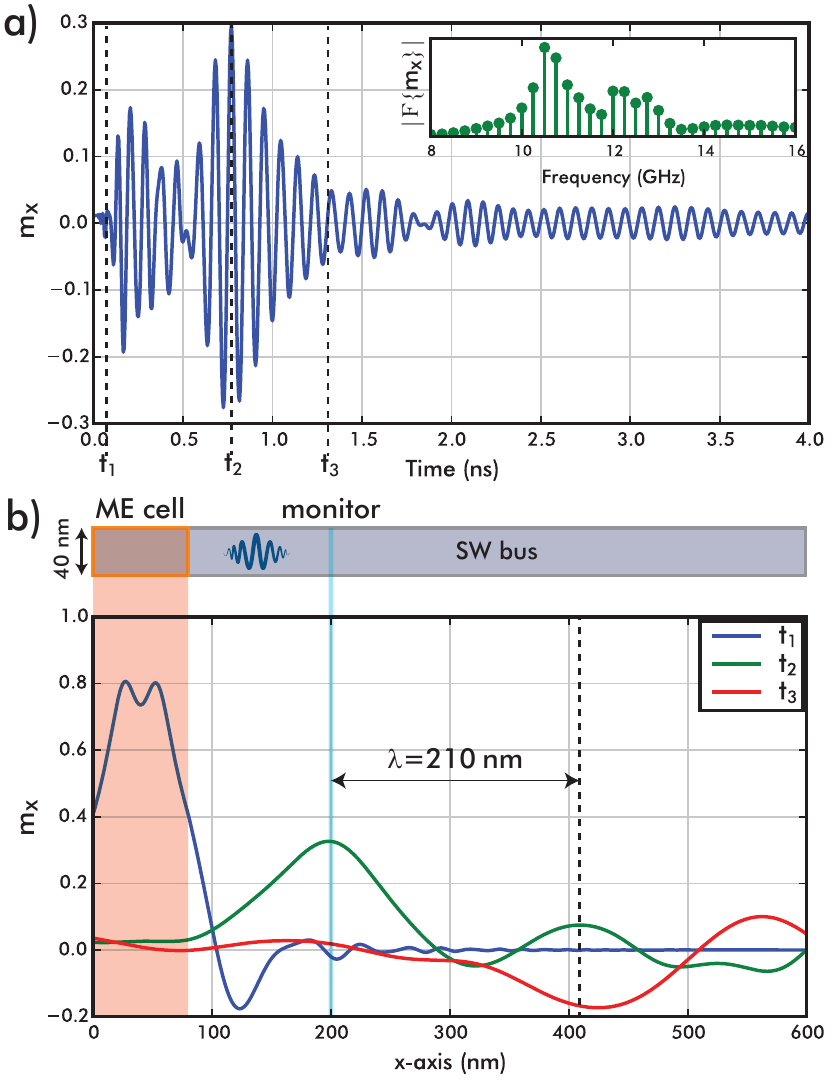}
	\caption{\label{fig:spectrum} Spin wave generated by ME cell. a) Temporal m$_{\textrm x}$ profile. Inset: Frequency components of propagated spin wave. b) Spatial m$_{\textrm x}$ profile of the spin wave bus at three different timepoints as denoted in a).}
\end{figure}

The temporal m$_{\textrm x}$ profile of the spin wave generated by an ME cell is shown in FIG. \ref{fig:spectrum}a. We observe that the spin wave created has a wave packet-like form, with multiple frequency components (as shown in the inset of FIG. \ref{fig:spectrum}a\cite{supp}) and duration shorter than 2\,ns. The structure simulated to generate the spin wave in FIG. \ref{fig:spectrum}a is depicted in FIG. \ref{fig:spectrum}b. An ME cell is activated and generates a spin wave that propagates along a spin wave bus. The magnetization dynamics are monitored after 120\,nm. FIG. \ref{fig:spectrum}b also shows the spatial m$_{\textrm x}$ profile at three different timepoints ($t_1$, $t_2$, $t_3$). At time $t_1$=0.065\,ns, the ME cell has not switched out-of-plane and the spin wave is not formed yet. At time $t_2$=0.77\,ns, the spin wave is formed and has propagated at least 120\,nm but is almost completely dispersed after $t_3$=1.3\,ns. Due to the complex nature of the spin wave, it's impossible to extract an accurate wavelength but from the m$_{\textrm x}$ profile at $t_2$, we can extract its wavelength at the largest amplitude is $\lambda$=210\,nm.

\begin{figure}[!ht]
	\includegraphics{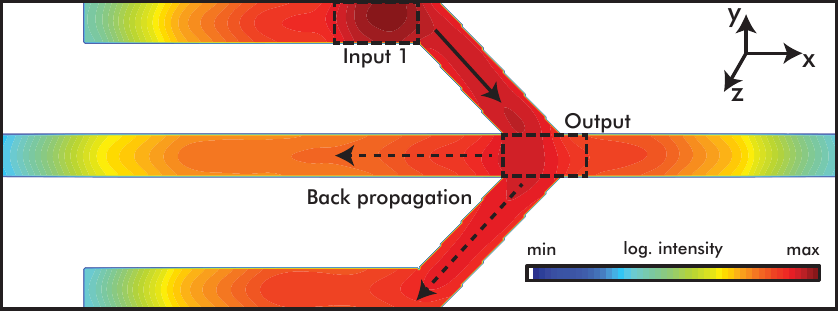}
	\caption{\label{fig:singleA} Spin wave amplitude transmission for single arm excitation of `Input 1', plotted in a logarithmic scale. Dashed arrows demonstrate the flow of back-propagated spin waves into the other input arms.}
\end{figure}

For the initial study of the spin wave majority gate's performance, we conducted single-arm excitation simulations and monitored the spin wave transmission in the complete structure. FIG. \ref{fig:singleA}, presents the spin wave amplitude (defined as  $\sqrt{{\textrm m}_{\textrm x}^2+{\textrm m}_{\textrm y}^2}$) averaged over time (i.e. 3\,ns) in logarithmic scale. The amplitude transmission from `Input 1' to `Output' is $\simeq$93\%, defined as the ratio of the average intensity of the output to the average intensity of the input. This efficient transmission is due to the nanoscale dimensions of the structure in combination with the low damping values of the materials assumed. The downside of the efficient transmission is that there is significant reflections and back-propagations (i.e. $\simeq$89\%, denoted by dashed arrows in FIG. \ref{fig:singleA}). This is due to the geometrical symmetry of the structure (unlike Klinger \textit{et al}\cite{klinger2}).

The back-propagations increase the complexity of the spin wave dynamics and interference but will not affect the states of ME cells that can be interconnected before the majority gate. The ME cell concept applied in this work ensures logical non-reciprocity\cite{waser} due to a three-phase clocking scheme.\cite{sourav1}

In order to have a functional spin wave majority gate, we need to ensure: (a) the input ME cells switch from in-plane to out-of-plane correctly and in a similar fashion; (b) the spin waves that arrive at the output region are as close to identical as possible (unbiased inputs); (c) that the output ME cell's detection operation is launched at the appropriate timepoint. The first requirement is satisfied since, when designing structure, we used the analytical expressions in Engel-Herbert \textit{et al}\cite{engel} and Kani \textit{et al}\cite{kani} to calculate the minimum arm spacing that also minimizes their dipolar coupling. This coupling would impede the ME cells to completely switch out-of-plane, thus not work properly. The minimum spacing of the arms is 56\,nm and is verified by simulations. To investigate the second requirement, we study the input signals by the means of the out-of-plane angle ($\theta$) as the angle between magnetization ($\textbf{M}$) and \zax{}.

\begin{figure}[!ht]
	\includegraphics{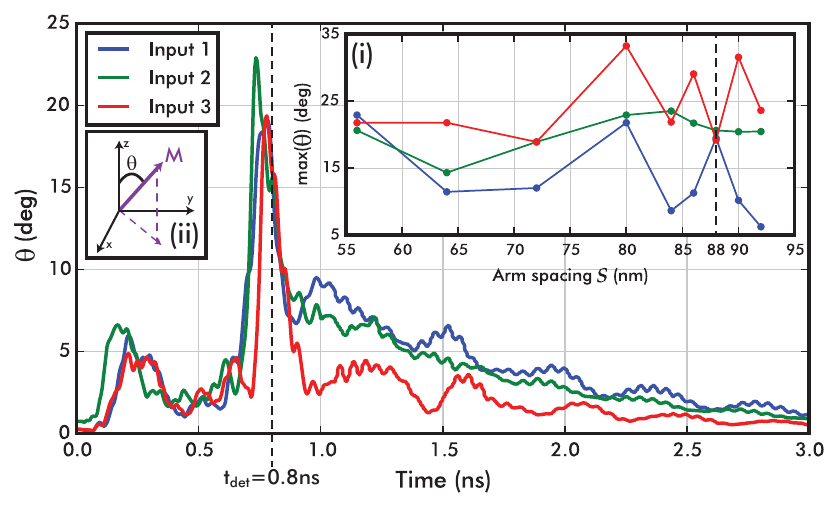}
	\caption{\label{fig:thetas} Average out-of-plane angle $\theta$ of the output magnetization when excited by individual single arm excitations in a structure with $S$=88\,nm. Based on $\theta$ for each spin wave signal, we select detection timepoint at $t_{det}$=0.8\,ns. Inset (i): Maximum $\theta$ of the output magnetization when excited by individual single arm excitations, shows that the selection of $S$=88\,nm as the arm spacing is the best one for the explored values. Inset (ii): definition of $\theta$.}
\end{figure}

The fork-like structure we employ has a mirror symmetry. However, the signals created by `Input 1' and `Input 3' do not follow that symmetry. The spin wave propagation and dispersion depends on the shape anisotropy variation that the $S$ parameter induces. This dependence is non-linear as demonstrated in inset (i) of FIG. \ref{fig:thetas}, where the maximum out-of-plane angle of the output magnetization (for each single arm excitation) is plotted over different values of $S$. FIG. \ref{fig:thetas}(i) shows that, by changing the geometry of the majority gate structure, the spin wave behavior changes. This means that, for each spacing value selected, the structure would have to be fine-tuned (in terms of material parameters and input ME cell positioning) to operate correctly. The latter hinders the robustness of the current geometry and needs to be evaluated further, including different geometry options. However, an accurate robustness evaluation is considered outside the scope of this work. We note that the spacing value $S$ where all three input signals have the most similar contributions to the output $\theta$ angle is at $S$=88\,nm. Hence these values were selected for a functional majority gate as they lead towards satisfying the second aforementioned requirement of unbiased inputs.

To further optimize the performance of the majority gate, through more micromagnetic simulations, we have defined the length of the spin wave bus that connects `Input 2' to `Output' at 92\,nm and a slightly increased damping of $\alpha$=0.016. Such local engineering of magnetic damping has been extensively studied\cite{king} and it could be implemented in the spin wave bus by controlled ion bean irradiation. This method ensures the PMA could be preserved whereas the magnetic damping diminishes due to increase surface roughness. With this configuration the requirement of the unbiased inputs is satisfied, as FIG. \ref{fig:thetas} shows that the spin wave signals from each input have almost identical contribution to the output magnetization.

The third requirement is satisfied by the detection timepoint of $t_{det}$=0.8\,ns, extracted from FIG. \ref{fig:thetas} where all three spin wave signals induce equal out-of-plane angle $\theta$. To verify the operation of the majority gate we need to excite all three inputs simultaneously and monitor the detected result. We define the logic `0' of the majority gate as the spin wave generated by an ME cell initially set along +\xax{} (m$_{\textrm{x}}$=1) and the logic `1' as the as the spin wave generated by an ME cell set along -\xax{} (m$_{\textrm{x}}$=-1). This definition is arbitrary.

\begin{figure*}[!ht]
	\includegraphics{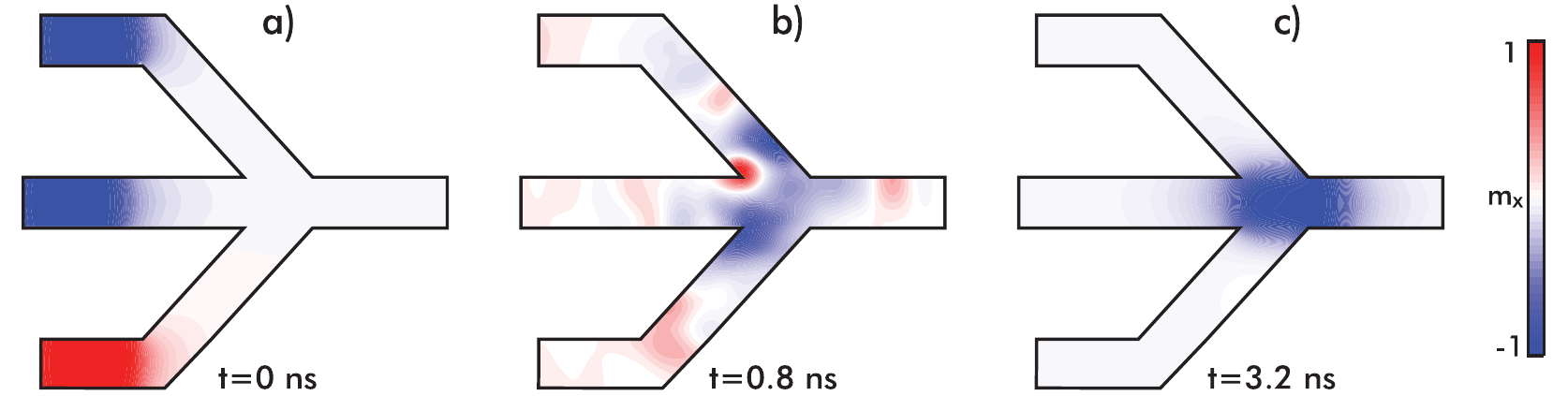}
	\caption{\label{fig:110snaps} Spatial profile of m$_{\textrm x}$ magnetization of the majority gate at different timepoints of operation. a) At t=0\,ns, the inputs are set to `110'. b) At t=0.8\,ns before the detection of the output ME cell is enabled, most of the magnetization oscillations are centered around the merging/output region. c) At t=3.2\,ns the output magnetization is stabilized to its non-volatile state `1' correctly detecting the majority result.}
\end{figure*}

FIG. \ref{fig:110snaps} illustrates an example operation of the spin wave majority gate, where the input are set to `110' (FIG. \ref{fig:110snaps}a). After the three inputs are activated, the generated spin waves propagate towards the output and interfere. At time t=0.8\,ns (FIG. \ref{fig:110snaps}b), the detection is enabled which results in the output ME cell to stabilize at the correct majority result `1' (m$_{\textrm{x}}$=-1 - FIG. \ref{fig:110snaps}c).

\begin{figure}[!ht]
	\includegraphics{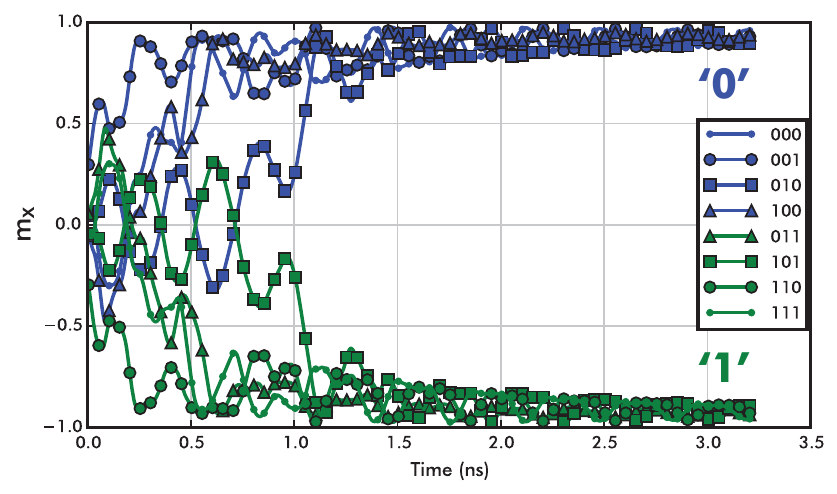}
	\caption{\label{fig:falling} Average m$_{\textrm x}$ of the output for all possible input combinations resulting in the correct majority computation.}
\end{figure}

Finally, to verify the complete logic behavior of the spin wave majority gate we simulate all possible input states. The results of these simulations are summarized in FIG. \ref{fig:falling}, where we observe that all inputs that have majority of `0' set the output ME cell magnetization along +\xax{} and all inputs that have majority of `1' set the output ME cell magnetization along -\xax{}. This proves the operation of the proposed design. Another interesting fact depicted in FIG. \ref{fig:falling} is that the output magnetization switching behavior is symmetric for symmetric inputs (e.g. for inputs `010' and `101'), which enhances the validity of the design as one that enables symmetrical and unbiased inputs.

The choice of as spin wave generators and detectors is not limited to ME cells, other effects such as Voltage-Controlled Magnetic Anisotropy (VCMA)\cite{vcma} could be used. However, the fact that the proposed majority gate utilizes the ME cell concept,\cite{sourav1} not only makes it non-volatile (characteristic of critical importance for low-energy applications) but also it provides the necessary means for cascading. Having detected and stored the majority result, the output ME cell could be easily triggered and generate the corresponding spin wave which will be detected by a cascaded ME cell interconnected with the spin wave bus. Additionally, having an ME cell operating voltage of 0.1\,V, results in an ultra-low intrinsic energy dissipation per ME cell of 4.5\,aJ.\cite{sourav3}

In conclusion, a fully functional, nanoscale, symmetric, non-volatile spin wave majority gate design utilizing ME cells as inputs and outputs, has been presented. The design was optimized for the correct detection of the majority result, without being disturbed by parasitic spin wave reflections and back propagations. The feature size of the design is 40\,nm and has a total area of 0.074\,$\mu$m$^2$, making it the smallest reported majority spin wave design to be functionally verified. Also, the proposed design operates in a $\sim$3\,ns timeframe which is fast compared to other spin-based technologies.\cite{nikonov} Finally, the combination of the proposed majority gate along with the ME cell inverter\cite{sourav1} and majority-based logic synthesis,\cite{amaru} can enable integrated circuit possibilities that exhibit ultra low-energy and small area characteristics.

\providecommand{\noopsort}[1]{}\providecommand{\singleletter}[1]{#1}%

\end{document}